\documentclass[12pt, a4paper]{article}
\pdfoutput=1

\usepackage{amsmath}
\usepackage{amsfonts}
\usepackage{amssymb}
\usepackage{graphicx, rotating}
\usepackage{epsfig}
\usepackage{array}
\usepackage{latexsym}
\usepackage{graphicx}
\usepackage{xcolor}
\usepackage{amsmath,bm,amssymb}
\usepackage{cite}
\usepackage{slashed}
\usepackage{hyperref}
\usepackage[normalem]{ulem}
\usepackage[utf8]{inputenc}
\usepackage[export]{adjustbox}
\usepackage{multirow, verbatim}

\newcommand{\be}{\begin{equation}}
\newcommand{\ee}{\end{equation}}
\newcommand{\bea}{\begin{eqnarray}}
\newcommand{\eea}{\end{eqnarray}}
\newcommand{\nn}{\nonumber}

\newcommand{\aNEW}{{\alpha_{\bar \theta}}}
\newcommand{\kNEW}{{\kappa_{\bar \theta}}}
\newcommand{\Tcr}{T_{\rm cr}}

\newcommand{\wplus}{w_+}

\newcommand{\ekin}{\rho_{\rm fl}}

\begin{document}

\begin{titlepage}

\begin{flushright}
\small
DESY 20-064
\end{flushright}
\vspace{.3in}

\begin{center}
{\Large\bf
Model-independent energy budget of \\
cosmological first-order phase transitions} \\ 
\vskip 0.2 cm
{\large\bf
A sound argument to go beyond the bag model}\\
\bigskip\color{black}
\vspace{1cm}{
  {\large
Felix~Giese,
Thomas Konstandin,
Jorinde~van~de~Vis
}}

{\small
DESY, Notkestra{\ss}e 85, D-22607 Hamburg, Germany
}

\bigskip

\begin{abstract}
We study the energy budget of a first-order cosmological phase transition, which is an important factor in the prediction of the resulting gravitational wave spectrum. Formerly, this analysis was based mostly on simplified models as for example the bag equation of state. Here, we present a model-independent approach that is exact up to the temperature dependence of the speed of sound in the broken phase. We find that the only relevant quantities that enter in the hydrodynamic analysis are the speed of sound in the broken phase and a linear combination of the energy and pressure differences between the two phases which we call pseudotrace (normalized to the enthalpy in the broken phase). The pseudotrace quantifies the strength of the phase transition and yields the conventional trace of the energy-momentum tensor for a relativistic plasma (with speed of sound squared of one third).

We study this approach in several realistic models of the phase transition and also provide a code snippet that can be used to determine the efficiency coefficient for a given phase transition strength and speed of sound. It turns out that our approach is accurate to the percent level for moderately strong phase transitions, while former approaches give at best the right order of magnitude.     
\end{abstract}

\end{center}

\end{titlepage}

\newpage
\section{Introduction}
\label{sec:Introduction}
Many models of physics beyond the Standard Model (SM) give rise to a first-order phase transition in the early universe. A first-order phase transition proceeds by the formation of bubbles in which the new, low-temperature phase is realized, expanding in a universe that is still in the old, high-temperature phase. Such a scenario is intriguing, firstly since it opens up the possibility that the baryon asymmetry was formed during this phase transition, for example during electroweak baryogenesis \cite{Cohen:1990py, Cohen:1990it,Nelson:1991ab,Cohen:1994ss, Morrissey:2012db, Konstandin:2013caa}. Secondly, the collision of bubbles can lead to a stochastic gravitational wave signal, that could be measured by future gravitational wave experiments. 

Gravitational waves can be an alternative probe of new physics to collider searches. For a first-order electroweak phase transition, the new physics should couple to the Higgs. In principle, this can lead to collider signatures, such as exotic Higgs decays, but also deviations of the coupling of the Higgs to other particles, depending on the nature of the new physics \cite{Davoudiasl:2012tu, Katz:2014bha}. Unfortunately, the LHC is not sensitive to a range of models that would give rise to a first-order phase transition. If the mass of the new particle is larger than half the Higgs mass, detection through direct decay is impossible. In so-called `nightmare scenarios' the new particle is a singlet under the SM gauge groups and has a $\mathbb{Z}_2$-symmetry which forbids Higgs-singlet mixing. The main collider signatures in such scenarios are direct production of the new particle, and deviations from the SM triple Higgs and Z-Higgs interactions, but testing this scenario requires a 100 TeV hadron collider or a next-generation lepton collider \cite{Noble:2007kk, Curtin:2014jma, Huang:2016cjm}.  
Colliders of any kind are even less sensitive to first-order phase transitions taking place in a dark sector. Yet, dark phase transitions could also play a role in the formation of the baryon asymmetry and lead to an observable gravitational wave signal \cite{Shelton:2010ta, Petraki:2011mv, Schwaller:2015tja, Hall:2019ank, Hall:2019rld}. Gravitational waves are a complementary probe and can give us information about particle physics in the very early universe, much above the energy scale of Big Bang Nucleosynthesis and the Cosmic Microwave Background.

The many detections of black hole and neutron star mergers by the LIGO/Virgo collaboration \cite{Abbott:2016blz, Abbott:2017oio, TheLIGOScientific:2017qsa} have demonstrated that gravitational waves are becoming a very effective probe of the (early) universe. Current generation gravitational wave experiments are designed to detect signals in the 10 Hz - 10 kHz range. The peak frequency of the gravitational wave signal of a cosmological first-order phase transition at electroweak temperatures lies in the mHz regime \cite{Grojean:2006bp}, and current experiments are therefore not suitable for observing cosmological phase transitions. However, the Laser Interferometer Space Antenna (LISA), that is planned to be launched in the next decade, would be optimally suited for detecting gravitational waves from a first-order phase transition that happened between the weak scale and the TeV scale. 

To determine whether a specific particle physics model gives rise to an observable gravitational wave signal, one needs to predict the gravitational wave spectrum. Without doing a full lattice study, the spectrum can be estimated as a function of the following ingredients:
\begin{itemize}
	\item The phase transition parameters, such as the percolation temperature and the phase transition duration. These parameters can be determined from the particle physics model analyzed at finite temperature.
	\item The bubble wall velocity, for which the equation of motion of the bubble wall needs to be solved in an out-of-equilibrium computation. In our work, the wall velocity is treated as an external parameter.
	\item The fraction of energy that is converted into fluid motion. This is the quantity of interest of our work and we will elaborate on its determination below.
	\item A numerical prefactor that is obtained from lattice simulations \cite{Hindmarsh:2015qta, Hindmarsh:2016lnk, Hindmarsh:2017gnf, Hindmarsh:2019phv}.
\end{itemize}
For more details, see e.g. \cite{Caprini:2015zlo, Caprini:2019egz}. 

The fraction of energy that is converted into fluid motion can be determined by solving the hydrodynamic equations of a single expanding bubble, as in \cite{LandauLifshitz, Kamionkowski:1993fg, Espinosa:2010hh}. Ref.~\cite{Espinosa:2010hh} uses the bag equation of state in their analysis, which describes the two phases as relativistic plasmas with a different number of relativistic degrees of freedom and a temperature-independent vacuum energy difference (the so-called bag constant). A fit of the kinetic energy fraction is given in terms of the phase transition strength $\alpha$ and wall velocity. In the literature, this fit is often used to easily determine the kinetic energy fraction, for a broad range of particle physics models. A problem with this approach is that the generalization of the phase transition strength to a different equation of state is not clear. Furthermore, the bag equation of state only allows for a speed of sound of $c_s^2 = 1/3$, corresponding to a relativistic plasma, whereas a realistic particle physics model might deviate from this value (especially in the broken phase).

The goal of this work is to generalize the study of Ref.~\cite{Espinosa:2010hh} to more realistic particle physics models, focusing on detonations. 
We will provide a model-independent analysis of the hydrodynamic fluid profile. Our main result is that for a temperature-independent speed of sound in the broken phase, only two parameters enter this analysis. The first one is the speed of sound in the broken phase, while the second is 
a linear combination of pressure and energy differences between the two phases (and the wall velocity) called $\aNEW$. 
Our new $\aNEW$ can be fully determined from the phase transition parameters, without solving the matching equations at the bubble wall. We demonstrate that, to good accuracy, the kinetic energy fraction of detonations is then also only a function of the above three parameters. We will benchmark our approach against the most common approximations of the kinetic energy fraction in the literature. 

The outline of this paper is as follows. In sec.~\ref{sec:HydroReview} we briefly review how to solve the hydrodynamic equations and determine the kinetic energy fraction in a concrete model. We discuss the efficiency of gravitational wave production in the bag model, and how this result is generalized in the literature to other models in sec.~\ref{sec:bag}. We introduce our model-independent generalization of the phase transition strength in sec.~\ref{sec:NewAlpha}. In sec.~\ref{sec:Nu} we compute the kinetic energy fraction in a simple model~ \cite{Leitao:2014pda} that allows for a deviation in the speed of sound $c_s^2 \neq 1/3$ and we demonstrate how our newly defined $\aNEW$ removes all further model dependence in the computation of the kinetic energy fraction. In sec.~\ref{sec:models} we compute the kinetic energy fraction for two more realistic models: one SM-like and one with a two-step phase transition. We do the full numerical computation and compare to five approximation schemes. We summarize and conclude in sec.~\ref{sec:summary}. Appendix~\ref{app:kappa} gives a Python code that can be handily used to determine the kinetic energy fraction for an arbitrary particle physics model.

\section{Bulk kinetic energy of the fluid}
\label{sec:HydroReview}

In this section we quickly review how to determine the bulk kinetic energy of the fluid for a single bubble. This analysis was first put forward in \cite{LandauLifshitz, Kamionkowski:1993fg, KurkiSuonio:1995pp, Espinosa:2010hh} and fitting functions to the so-called efficiency coefficient $\kappa$ have been presented in \cite{Espinosa:2010hh} for the bag equation of state. 

The hydrodynamics in the plasma is described in terms of the energy density $e$, the pressure density $p$, and the enthalpy $w$. The pressure is determined by the free energy
\be
p = - \mathcal{F},
\ee
while the energy $e$ and enthalpy $w$ are related in the following way
\be
e = T \frac{\partial p}{\partial T} - p\, , \qquad w = T \frac{\partial p}{\partial T} = e+p\, .
\ee
The free energy $\mathcal{F}$ is the finite temperature effective potential and can be determined with the established methods \cite{Dolan:1973qd,Weinberg:1974hy,Landsman:1986uw, Quiros:1994dr, Kapusta:2006pm} in every specific particle physics model. The energy momentum tensor of the fluid is then given by
\be
\label{eq:Tmunu}
T^{\mu\nu} = u^\mu u^\nu w + g^{\mu\nu} \, p \, ,
\ee
where $u^\mu$ denotes the fluid velocity four-vector and $g$ is the Minkowski metric. 

Ultimately, we are interested in the power spectrum of the stochastic gravitational waves produced by 
the motion of the fluid. On general grounds one expects the relation
\be
\Omega_{GW} \propto \left( \frac{\ekin}{e_+} \right)^2 \, ,
\label{eq:OmegaGW}
\ee
where $\Omega_{GW}$ denotes the energy fraction in gravitational wave radiation, $e_+$ denotes the energy density of the plasma before the phase transition and $\ekin$ is the kinetic energy density in the fluid~\cite{Huber:2008hg} (if turbulence develops in less than a Hubble time, this relation can be modified~\cite{Hindmarsh:2017gnf}). 

For a single bubble during the phase transition, the kinetic energy results from the integral over the trace of the spatial components of the energy-momentum tensor~(\ref{eq:Tmunu}) ($\xi_w$ is the wall velocity)
\be
\label{eq:K}
K \equiv \frac{\ekin}{e_+} \, , \quad
\ekin =   \frac{3}{\xi^3_w} \int d\xi \, \xi^2 \, v^2 \gamma^2\, w \, .
\ee
In this equation, the enthalpy $w(\xi)$, the fluid velocity $v(\xi)$ (in the rest frame of the bubble) and the Lorentz factor $\gamma(\xi) = 1/\sqrt{1-v(\xi)^2}$ are self-similar and only depend on  the ratio $\xi = r/t$, with $r$ the distance to the center of the bubble, and  $t$ the time since nucleation. That $\Omega_{GW}$ is indeed proportional to $K^2$ is very explicit in simplified simulations as \cite{Kamionkowski:1993fg, Huber:2008hg, Jinno:2017fby, Konstandin:2017sat, Jinno:2019jhi}.
For the full-fledged hydrodynamic simulations of the phase transition, this is not obvious and this proportionality might be slightly modified. We take the relationship~(\ref{eq:OmegaGW}) for granted and the main aim of the present work is to determine $K = \ekin/e_+$ as accurately as possible in a model-independent way.

In order to obtain the fluid and temperature profiles $v(\xi)$ and $T(\xi)$, the hydrodynamic equations are solved for one bubble that nucleates and expands. The plasma is then described by two different phases that fill the inside/outside of the bubble and accordingly two different free energies. 

Conservation of the energy momentum tensor across the phase transition front yields a set of equations relating the 
energy-momentum tensor in the two phases. The key feature of these equations is that the different free energies in the two phases imply a change in temperature $T_{\pm}$ and fluid velocity $v_{\pm}$ across the wall interface. 
Note that $v_\pm$ are defined in the rest frame of the bubble wall.
The matching equations read~\cite{LandauLifshitz}
\bea
\frac{v_+}{v_-}&=&\frac{e_b(T_-)+p_s (T_+)}{e_s (T_+)+p_b (T_-)} \, , \\
    v_+ v_-&=& \frac{p_s(T_+)-p_b(T_-)}{e_s(T_+)-e_b(T_-)} \, , 
    \label{eq:matching}
\eea
where the subscripts $s$ and $b$ indicate the symmetric and broken phases. At first sight, the matching equations depend on the pressure and energy in both phases, but since the velocities are dimensionless, only three ratios can be relevant. It is perhaps surprising, that in fact only two quantities enter, which becomes transparent by rewriting
\bea
\frac{v_+}{v_-}&=&\frac{1 - \Delta e/w_+}{1 - \Delta p/w_+} \, , \label{eq:matching2a}\\
    v_+ v_-&=& \frac{\Delta p/w_+}{\Delta e/w_+} \, ,
    \label{eq:matching2}
\eea
where we defined $w_+ \equiv w_s(T_+)$ and 
\be
\Delta p \equiv p_s(T_+)-p_b(T_-) \, , \quad \Delta e \equiv e_s(T_+)-e_b(T_-) \, . \label{eq:DeltaPDeltaE}
\ee
In a detonation, the temperature in front of the wall is the nucleation temperature of the phase transition and the fluid velocity in front of the wall $v_+$ coincides with the wall velocity $\xi_w$ (that has to be determined by a friction calculation as done for example in \cite{Moore:1995ua, Moore:1995si, Konstandin:2014zta, Dorsch:2018pat}). Hence these two equations determine the quantities $v_-$ and $T_-$ behind the wall in terms of $v_+$ and $T_+$. 

Finding the velocity profile of the plasma away from the bubble wall is simplified by the fact that the solution is self-similar and only depends on $\xi = r/t$, where $r$ is the radial coordinate of the bubble and $t$ the time since nucleation. Energy-momentum conservation then leads to the two equations
\bea
(\xi - v) \frac{\partial_\xi e}{w} &=& 2 \frac{v}{\xi} + \gamma^2 (1 - \xi v)\partial_\xi v \, ,\\ 
(1 - \xi v) \frac{\partial_\xi p}{w} &=& \gamma^2 (\xi - v) \partial_\xi v \, .
\eea
For a general equation of state, these two equations can be read as coupled differential equations for $\partial_\xi v$ and $\partial_\xi T$. In case the speed of sound 
\be
c_s^2 = dp/de \, ,
\ee
is constant, these two equations can be decoupled into
\be
\label{eq:dvdxi}
2\frac{v}{\xi}=\gamma^2\left(1-v\xi\right)\left[\frac{\mu^2}{c_s^2}-1\right] \partial_\xi v \, ,
\ee
and
\be
\frac{\partial_v w}{w} = (1/c_s^2 + 1) \gamma^2 \mu \, ,
\ee
with $\mu$ the boosted fluid velocity 
\be
\mu(\xi,v)=\frac{\xi-v}{1-\xi v}.
\ee
The solutions to these equations can be used in~(\ref{eq:K}) to determine $K =\ekin/e_+$.

Before we further discuss this system of equations let us comment on energy-momentum conservation in its
integrated form. Integration of $T^{00}$ over the volume of the bubble yields the relation
\be
\int d\xi \, \xi^2 (\gamma^2 w - p) = \frac13 \xi_w^3 e_+ \, ,
\ee
or equivalently 
\be
\label{eq:energyconservation}
\ekin = \frac{3}{\xi_w^3} \int d\xi \, \xi^2 (e_+ - e)\, .
\ee
The interpretation of this relation is that $K = \ekin/e_+$ quantifies the fraction of energy that is converted into fluid motion~\footnote{Notice that there is a typo in the last relation of (32) in \cite{Espinosa:2010hh}.} and hence $0<K<1$. 

    \section{Efficiency and the bag equation of state\label{sec:bag}}

In this section we discuss the efficiency of gravitational wave production in the bag equation of state in order to make contact with the existing literature. We also discuss how the results of the efficiency in the bag model are applied to other models in the literature.

First, let us consider the bag equation of state as done in \cite{Espinosa:2010hh} 
\be
\label{eq:bagP}
p_s = \frac13 a_+ T^4 - \epsilon, \quad p_b = \frac13 a_- T^4 \, .
\ee
Accordingly
\be
e_s = a_+ T^4 + \epsilon, \quad e_b = a_- T^4 \, ,
\ee
and
\be
w_s = \frac43 a_+ T^4, \quad w_b = \frac43 a_- T^4 \, .
\ee
The critical temperature (where the pressures are degenerate) is hence given by the relation $T_{\rm cr}^4 = 3\epsilon/(a_+ - a_-)$
and $a_+ > a_-$ is implied. In the bag equation of state the speed of sound in both phases is $c_s^2=dp/de=1/3$ whereas in general the speed of sound is a function of temperature and depends also on the phase.

For the bag equation of state, the matching equations~(\ref{eq:matching}) simplify to
\be
\frac{v_+}{v_-} = \frac{3+(1-3\alpha_\epsilon) r}{1+3(1+\alpha_\epsilon) r} \, , \quad
v_+ v_- = \frac{1-(1-3\alpha_\epsilon) r}{3-3(1+\alpha_\epsilon) r} \, ,
\ee
with
\be
\alpha_\epsilon = \frac{\epsilon}{a_+ T_+^4} = \frac{4 \epsilon}{3 w_+} \, , \quad
r = \frac{a_+ T_+^4}{a_- T_-^4} = \frac{w_+}{w_-}\, .
\ee
Since the fluid kinetic energy $\ekin$ is sourced by the bag constant $\epsilon$, this motivates the definition of the 
`efficiency coefficient' $\kappa_\epsilon$
\be
\kappa_\epsilon \equiv \frac{\ekin}{\epsilon} \, ,
\ee
and
\be
K = \frac{\ekin}{e_+} = \frac{\alpha_\epsilon \, \kappa_\epsilon}{\alpha_\epsilon + 1} \, .
\ee
In the bag model $\kappa_\epsilon$ is only a function of the wall velocity $\xi_w$ and the strength parameter $\alpha_\epsilon$.

In other models for the phase transition, the above results can be potentially reused if there is a way to infer a bag constant $\epsilon$ that will give a reasonable result. Notice that while $\kappa_\epsilon$ is called `efficiency coefficient' in the literature, one could also call  $K$ an `efficiency coefficient' owing to the relation~(\ref{eq:energyconservation}). In any case, we are mostly concerned with the quantity $K = \ekin/e_+$ that sometimes is also called energy fraction.

In the literature, the most common mapping to $\alpha_\epsilon$ is to either use the pressure difference 
\be
\alpha_{p} = -\frac{4 Dp}{3 w_s(T_+)} \equiv  -\frac{4(p_s(T_+) - p_b(T_+))}{3 w_s(T_+)} \, ,
\ee
or the energy difference
\be
\alpha_{e} = \frac{4 De}{3 w_s(T_+)} \equiv  \frac{4(e_s(T_+) - e_b(T_+))}{3 w_s(T_+)} \, .
\ee
All these quantities are evaluated at the temperature $T_+$, since using $T_-$ would require to solve the matching equations in the first place, which one would like to avoid. Note that we introduced the notation with a $D$ if both quantities are evaluated at $T_+$ and $\Delta$ when the two quantities are evaluated at $T_+$ and $T_-$, respectively. 

The rationale behind using the pressure difference is that for a vanishing pressure, the wall becomes static and hence $K$ should become small. This is true but somewhat misguided since in the limit of vanishing pressure difference the wall velocity will vanish (which in turn enforces $K \to 0$). However, when the wall velocity  is imposed as external input there is no physical reason for $K$ to vanish even in case the pressure difference is zero (imagine a system where the wall is pushed by some external force, the fluid still would establish some hydrodynamic solution and produce GWs). Likewise, the energy difference is motivated by the relation~(\ref{eq:energyconservation}) and the understanding that the energy difference has to fuel the fluid kinetic motion. 
As we will see later, also this fails to provide a good mapping.

Finally, the bag constant can be mapped using the trace of the energy momentum tensor ($\theta = g_{\mu\nu} T^{\mu\nu}$) which has been advocated in \cite{Espinosa:2010hh, Hindmarsh:2015qta, Hindmarsh:2017gnf, Hindmarsh:2019phv}
\be
D\theta = De - 3 Dp \, , \quad 
\alpha_{\theta} = \frac{D\theta}{3 w_s(T_+)} \, .
\ee
In the bag model $\alpha_\theta = \alpha_\epsilon$ such that $\alpha_\theta$ is a faithful generalization of $\alpha_\epsilon$. 
We will see that the mapping using $\alpha_{\theta}$ works reasonably well, in particular when the speed of sound is close to $c_s^2 \simeq 1/3$.

For strong phase transitions, $T_+ \ll T_{\rm cr}$, the thermal part of the free energy is irrelevant and $\alpha_{p} \simeq \alpha_{e} \simeq \alpha_\theta$. However, for weak phase transitions $T_+ \simeq T_{\rm cr}$, one finds $\alpha_{p}\simeq 0$ while $\alpha_{e}\simeq 4\alpha_\theta$. In general, due to $Dp<0$ and $\frac{d}{dT}(Dp)>0$ one has
\be
\label{eq:chain}
\alpha_{p} < \alpha_\theta < \alpha_{e} \, .
\ee
In the next section we will present a model-independent analysis and a first assessment on which of the three choices performs best. 
    
    \section{Model-independent matching equations}\label{sec:NewAlpha}

We want to find a generalization of $\alpha_\epsilon$ that reproduces the correct kinetic energy fraction independently of the details of the model. This $\alpha$ can be a function of the model parameters and the nucleation temperature, which in the case of detonations is $T_+$. Dependence on the temperature in the broken phase, $T_-$, is undesirable, since this quantity can only be obtained by solving the matching equations at the bubble wall. 

We therefore expand our thermodynamic quantities around the symmetric phase to obtain their values in the broken phase. This will allow us to eliminate $T_-$.
To be specific, we write
\bea
\Delta p &=& p_s(T_+) - p_b(T_-) \nn \\
&=& [p_s(T_+) - p_b(T_+)] + [p_b(T_+)- p_b(T_-)] \nn  \\
&\equiv& \quad\quad Dp \qquad\qquad \, + \quad\quad\delta p \, ,
\eea
and likewise for energy and enthalpy.

At this point, in order to make progress we have to introduce an additional assumption. Namely, that the phase transition is not too strong. Under this assumption, $T_+ \simeq T_-$ and the quantities $\delta p$ and $\delta e$ are related via the speed of sound
\be
\frac{\delta p}{\delta e} \simeq \left. \frac{dp_b/dT}{de_b/dT} \right|_{T_+}  \equiv c_s^2\, .
\label{eq:def:sos}
\ee
Please note that this is the speed of sound defined using the thermodynamic potentials in the broken phase.
The second matching equation then reads
\be
\delta p \, (1- v_+v_-/c_s^2) \simeq v_+ v_- De  - Dp \, ,
\ee
and the ratio of the velocities is then given by:
\be
\frac{v_+}{v_-} \simeq \frac{w_+(v_+v_-/c_s^2-1) + (De - Dp/c_s^2)}{w_+(v_+v_-/c_s^2-1) + v_+v_-(De - Dp/c_s^2)} \, .
\label{eq:vovNEW}
\ee
This motivates the key definitions of the \emph{pseudotrace} $\bar\theta$ and the corresponding strength parameter. So we define
\be
\boxed{
\bar \theta \equiv e - p/c_s^2 \, , \quad
\aNEW \equiv \frac{D\bar \theta}{3w_+} \, , 
}\, \label{eq:def:pt} 
\ee
such that 
\be
\frac{v_+}{v_-} \simeq \frac{(v_+v_-/c_s^2-1) + 3\aNEW}{(v_+v_-/c_s^2-1) + 3 v_+v_- \aNEW} \, .
\label{eq:omatchnew}
\ee
Let us reiterate that we assumed a weak phase transition, $T_- \simeq T_+$ in the approximation~(\ref{eq:def:sos}), but otherwise presented a model-independent analysis. 
As we will see later, in some models the relation~(\ref{eq:def:sos}) is exact and the matching equations are determined solely by
$\aNEW$ and $c_s^2$. In other words, the accuracy of~(\ref{eq:omatchnew}) hinges on the question how temperature independent the speed of sound in the broken phase is. 

How surprising is the result of equation~(\ref{eq:vovNEW})? In the limit of weak phase transitions, $Dp, De \ll w_+$, only a linear combination of $De$ and $Dp$ enters at leading order in the matching equations but it is by no means automatic that this linear combination does not depend on $v_+ v_- \simeq \xi_w^2$. 
Even more remarkably, the same is true for strong phase transitions and $\aNEW$ and $c_s^2$ are the only relevant quantities that enter in the matching equations as long as the speed of sound is temperature-independent. 

Note that if the matching only depends on $\aNEW$ and $c_s^2$, so do $\Delta p/w_+$ and $\Delta e/w_+$ (but not $Dp/w_+$, $De/w_+$, $\delta p/w_+$ or $\delta e/w_+$). In turn this is also true for $\Delta w/w_+$ and, in case the temperature dependence of $c_s^2$ is negligible, this also holds for $\ekin/w_+$ and 
\be
\label{eq:kappa_ptr}
\kappa_{\bar \theta} = \frac{4\ekin}{D\bar \theta} \, ,
\ee
which generalizes $\kappa_\epsilon$ from the bag model. The same is not true for $K = \ekin/e_+$ since 
two models with the same $\aNEW$ do not necessarily have the same $e_+/w_+$. As a trivial example, notice that the hydrodynamic analysis does not depend on the cosmological constant, and it does not matter if the bag constant is attributed to the broken or the symmetric phase. This is expected since only gravity itself is sensitive to a cosmological constant. The gravitational spectrum on the other hand depends on the Hubble parameter which is why it is sensitive to the cosmological constant. But even two models with the same energy density at $T=0$ and the same $\aNEW$ can have a drastically different adiabatic index $\Gamma = w_+/e_+$ at the phase transition which is why $K$ cannot be a function of $\aNEW$ alone.

The main outcome of above analysis is that the efficiency $\kappa_{\bar \theta}$ mostly depends on the wall velocity $\xi_w = v_+$, the speed of sound in the broken phase $c_s^2$ and the new phase transition strength parameter $\alpha_{\bar \theta}$. As a corollary, in the case when the speed of sound is that of a relativistic plasma, $c_s^2 = 1/3$, the trace will quantify the strength of the phase transition properly and the fits to the efficiency coefficient in the bag model~\cite{Espinosa:2010hh} apply. Analogously to the last section, we find the chain of inequalities
\be
\label{eq:chainNEW}
\alpha_{p} < \frac{4 \, \aNEW}{1+1/c_s^2} < \alpha_\theta < \alpha_{e} \, .
\ee

In the remainder of the paper we present several numerical tests and also discuss a generalization of the bag model that allows for a varying speed of sound (the $\nu$-model). The efficiency coefficient in any other model can then be inferred by mapping to this model using the strength parameter based on the pseudotrace, $\alpha_{\bar \theta}$ and the speed of sound $c_s$. As we will see, whenever the speed of sound departs significantly from $c_s^2=1/3$, the energy fraction in the new parameterization using $\aNEW$ will be much more accurate and the fits from \cite{Espinosa:2010hh} do not apply.

\section{Bulk motion in the $\nu$-model}\label{sec:Nu}

%
\begin{figure}
	\begin{center}
		\includegraphics[width= 1.0\textwidth]{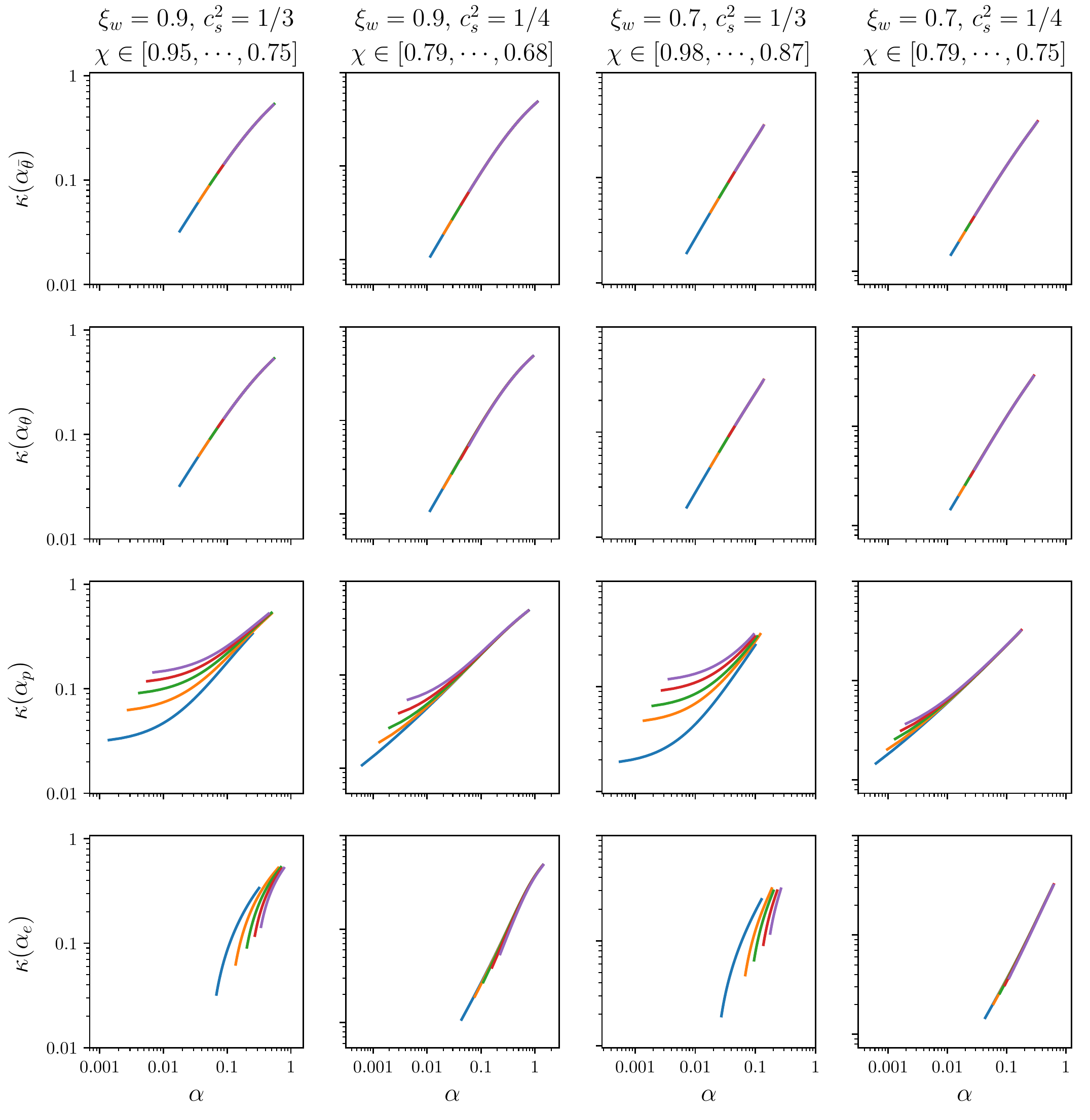}
		\caption{\small Efficiency $\kappa_{\bar \theta}$ for the $\nu$-model. The four rows show $\kappa_{\bar \theta}$ as a function of $\aNEW$, $\alpha_\theta$, $\alpha_p$ and $\alpha_e$. The four columns correspond to different values of $\xi_w$ and $c_s^2$. The plots show that only $\aNEW$ and potentially $\alpha_\theta$ can provide a model-independent assessment of the efficiency coefficient. The various colors are $\nu$-models with different parameters $\chi$.
			\label{fig:kappas4x4}}
	\end{center}
\end{figure}

In order to test the hypothesis from the last section, one needs to set up a model that allows for strong 
phase transitions and also for a significant change of the speed of sound in the broken phase. 
The simplest model that provides these features is the model already discussed in \cite{Leitao:2014pda} (which we will call $\nu$-model)
\bea
\label{eq:eos_nu}
p_s=\frac13 a_+ T^4 -\epsilon \, , \qquad  && 
e_s=a_+ T^4 + \epsilon\, ,  \nn \\ 
p_b=\frac13 a_- T^\nu \, , \qquad && 
e_b= \frac13 a_- (\nu-1) T^\nu \,,
\eea
where we have chosen a relativistic speed of sound in the symmetric phase. The free parameter $\nu$ can be eliminated in favor of the speed of sound in the broken phase using 
\be
\label{eq:cs2}
\nu=1+\frac{1}{c^2_s} \, . 
\ee
The speed of sound in the $\nu$-model is temperature independent which in turn is also its defining property: A constant speed of sound automatically implies the equation of state given in (\ref{eq:eos_nu}). This will simplify the analysis in the $\nu$-model and allow for an analytic treatment of many relations.

One can easily see that the $\nu$-model realizes the bag equations of state when the speed of sound in the broken phase equals the value in the bag model $c_s^2=1/3$ and $\nu=4$.
In the $\nu$-model, $\aNEW$ is given by 
\be
\label{eq:anew_numodel}
\aNEW = \frac{1}{12} \left(4-\nu + \frac{3\epsilon \nu}{a_+ T_+^4} \right).
\ee
Using one of our matching equations to eliminate $T_-$, we get the following relation between $v_+$ and $v_-$ 
\be
\frac{v_+}{v_-} = \frac{(v_+v_-(\nu-1)-1) + 3\aNEW}{(v_+v_-(\nu-1)-1) + 3 v_+v_- \aNEW} \, ,
\ee
which allows us to determine $v_-$ as a function of $c_s$, $v_+$ and $\aNEW$ only. In contrast to~(\ref{eq:omatchnew}), this is not an approximation but an exact relation in the $\nu$-model, owing to the fact that the speed of sound is constant in both phases. The corresponding efficiency coefficient $\kappa_{\bar \theta}$ will therefore also only depend on these quantities. Notice, that we have to impose that $\aNEW$ is positive, otherwise the matching equations do not allow for 
detonations~\footnote{The $\nu$-model also can have two phase transitions, from the broken phase to the symmetric one and back at a lower temperatures. We do not explicitly avoid this subset of models but only study the phase transition from the symmetric to the broken phase.}.

The model has overall five parameters that enter in the hydrodynamic analysis: $a_-$, $a_+$, $\epsilon$, $\nu$ and also the nucleation temperature $T_+$. The overall size of the thermodynamic potentials is irrelevant and only dimensionless combinations can enter the analysis. Hence, once the speed of sound $c_s^2$ and $\aNEW$ are fixed, one parameter remains free. One possible choice is to define as free parameter
\be
\chi = \frac{a_-}{a_+} \, T_{\rm cr}^{\nu-4} \, ,
\ee
one can then eliminate $\epsilon$ for the critical temperature
\be
\epsilon = \frac 1 3(a_+ T_{\rm cr}^4- a_- T_{\rm cr}^\nu) = \frac{1}{3} a_+ T_{\rm cr}^4 (1 - \chi)\, ,
\ee  
$\nu$ for the speed of sound using~(\ref{eq:cs2}) and the nucleation temperature $T_+$ for $\aNEW$ using~(\ref{eq:anew_numodel})
\be
(1-\chi) \left(\frac{T_{\rm cr}}{T_+}\right)^4 = \frac{12 \aNEW+(\nu-4)}{\nu} \, .
\ee
The individual values of $a_+$, $a_-$ and $T_{\rm cr}$ will only enter through the combination $\chi$ in the analysis. 
Furthermore, different values of $\chi$ will produce different values of $\alpha_\theta$ (in case $c_s^2 \not= 1/3$) and also different values of 
$\alpha_p$ and $\alpha_e$.

In Fig.~\ref{fig:kappas4x4} we show $\kappa_{\bar \theta}$ as defined in (\ref{eq:kappa_ptr}) for different $\nu$-models as a function of $\aNEW$, $\alpha_\theta$, $\alpha_p$ and $\alpha_e$ (in the four different rows). It is clearly visible that models with different $\chi$ coincide when $\kappa_{\bar \theta}$ is plotted as a function of $\aNEW$ or $\alpha_\theta$. This is not true for $\alpha_p$ or $\alpha_e$. Also as a function of $\alpha_\theta$, small deviations are present which are however not visible in this double-logarithmic plot. A more detailed comparison between $\aNEW$ and $\alpha_\theta$ in more realistic models is given in the next section. As a first conclusion, we remark that using $\alpha_p$ or $\alpha_e$ to quantify the strength of the phase transition will lead to an efficiency coefficient that is wrong by up to an order of magnitude. 

The reader might wonder why we have used $\kappa_{\bar \theta}$ in our comparison, and not for example $K$, the quantity that we are ultimately interested in. As explained above, the value of $K$ depends on $\chi$ via its dependence on $e_+$ and is therefore not a good quantity. Another model-independent choice would have been the ratio $\bar U^2 _f = \ekin/\wplus$. This ratio also would lead to the same conclusion while it does not refer explicitly to $\aNEW$ or $D\bar \theta$ in its definition (unlike $\kappa_{\bar \theta}$).

As we will show in the next section, the $\nu$-model is a useful tool in the determination of $K$ in more realistic models. The hydrodynamics of more complicated models can be mapped onto the simpler $\nu$-model via $\aNEW$ to very good accuracy. We denote the efficiency coefficient $\kappa_{\bar \theta}$ in the $\nu$-model as $\kappa_{\bar\theta}(\alpha_{ \bar \theta}, c_s)\rvert_\nu$.

\section{More realistic models}\label{sec:models}

In this section we study two more realistic models and discuss to which extent a good estimate of the efficiency coefficient can be obtained 
by mapping these models to the $\nu$-model using the speed of sound $c_s$ and the strength parameter $\aNEW$.

The first model is SM-like with a cubic term in the free energy coming from thermal effects. The thermodynamic potentials in the symmetric and broken phases are given via the free energy
\bea
{\cal F}(\phi,T) &=& -\frac{a_+}{3} T^4 + \lambda ( \phi^4 - 2 E \phi^3 T +\phi^2 (E^2 T_{\rm cr}^2 + c (T^2 - T_{\rm cr}^2)) ) \nn \\
&& \quad + \frac{\lambda}{4} (c-E^2)^2 T_{\rm cr}^4    \, .
\eea
such that $p_s = -{\cal F}(0,T)$ and $p_b = -{\cal F}(\phi_{\rm min},T)$ with 
\be
\phi_{\rm min} = \frac34 E \, T + 
\sqrt{T^2 \, (9E^2/8 - c)/2 - \Tcr^2 \, (E^2 - c)/2} \, .
\ee
The model has four relevant parameters: $3\lambda/a_+$, $E$, $T_+/T_{\rm cr}$ and $c$, which in turn determine $\aNEW$ and $c_s$. The last term in the free energy removes the cosmological constant at zero temperature in the broken phase. Symmetry breaking at low temperatures requires $c>E^2$. The barrier persists down to a temperature 
$T^2>T_{\rm cr}^2 (c-E^2)/(c-9E^2/8)$.

Table~\ref{tab:SM} shows our choices for four example models. The phase transitions are only weak to moderately strong which implies a speed of sound that is close to the value in a relativistic plasma.
\begin{table}
\begin{center}
\begin{tabular}{ | c  || c | c | c | c || c | c |}
\hline
Model & $3\lambda/a_+$ & $E$ & $c$ & $T_+/T_{\rm cr}$ & $\aNEW$ & $c_s^2$ \\
\hline
\hline
${\rm SM}_1$ & 10 & 0.3 & 0.2 & 0.9 & 0.0297 & 0.326 \\
\hline
${\rm SM}_2$ & 10 & 0.3 & 0.2 & 0.8 &  0.0498 & 0.331 \\
\hline
${\rm SM}_3$ & 3 & 0.3 & 0.2 & 0.9 & 0.00887 & 0.331 \\
\hline
${\rm SM}_4$ & 3 & 0.3 & 0.2 & 0.8 &  0.0149 & 0.333 \\
\hline
\end{tabular}
\caption{\small\label{tab:SM} Parameters of the SM-like models.}
\end{center}
\end{table}
\begin{table}
	\begin{center}
		\begin{tabular}{ | c  || c | c | c | c || c | c |}
			\hline
			Model & $b_-/(\sqrt{a_+}T^2_{\rm cr})$ & $c_-/\sqrt{a_+}$ & $c_+/\sqrt{a_+}$ & $T_+/T_{\rm cr}$ & $\aNEW$ & $c_s^2$ \\
			\hline
			\hline
			${\rm 2step}_1$ & $0.4/\sqrt{3}$ & $0.2/\sqrt{3}$ & $0.1/\sqrt{3}$ & 0.9 & 0.0156 & 0.312\\
			\hline
			${\rm 2step}_2$ & $0.4/\sqrt{3}$ & $0.2/\sqrt{3}$ & $0.1/\sqrt{3}$ & 0.7  & 0.0704 & 0.297\\
			\hline
			${\rm 2step}_3$ & $0.5/\sqrt{3}$ & $0.4/\sqrt{3}$ & $0.2/\sqrt{3}$ & 0.9  & 0.0254 & 0.282\\
			\hline
			${\rm 2step}_4$ & $0.5/\sqrt{3}$ & $0.4/\sqrt{3}$ & $0.2/\sqrt{3}$ & 0.7 & 0.159 & 0.245\\
			\hline
		\end{tabular}
		\caption{\small\label{tab:2step} Parameters of the models with two-step phase transition. }
	\end{center}
\end{table}

The second model we study is a model with a two-step phase transition~\cite{Espinosa:2011ax}. The model has two scalar fields that break for example the electroweak symmetry and a $\mathbb{Z}_2$ symmetry. Although some symmetry is broken in both phases, we will still denote the phase that the field tunnels through first as `symmetric' and the second phase as `broken'. This time we neglect the cubic term. The pressure in the two phases can then be brought to the form
\bea
p_s(T) &=& \frac13 a_+ T^4 + (b_+ - c_+ T^2)^2 - b_-^2 \, , \\
p_b(T) &=& \frac13 a_+ T^4 + (b_- - c_- T^2)^2 - b_-^2 \, ,
\eea
where we have subtracted the same cosmological constant in both phases. We can express one of the parameters using the critical temperature via $b_- - b_+ = T_{\rm cr}^2(c_- - c_+)$. Again the model has four relevant parameters, for example: $b_-/(\sqrt{a_+}T^2_{\rm cr})$, $c_-/\sqrt{a_+}$, $c_+/\sqrt{a_+}$ and $T_+/T_{\rm cr}$. Notice that by construction, the speed of sound in the symmetric phase also deviates from $c_s^2 = 1/3$. However, this is not relevant in our approach. In fact, one can generalize the $\nu$-model by introducing another free parameter for the speed of sound in the symmetric phase and still, $\aNEW$ and the speed of sound in the broken phase would be the only relevant input. The parameter $\aNEW$ will depend on the speed of sound in the symmetric phase in this case though. 

Table~\ref{tab:2step} shows our choices for four example models. The phase transitions are moderately strong to strong and the speed of sound in some of the models is below $c_s^2<1/4$.

With these eight models at hand, we compare different approximation schemes to determine $K$. The first is the full numerical evaluation of the model without any further approximations. The second matches the model to the $\nu$-model with the same $\aNEW$ and $c_s$. This is the method we would advocate for phenomenological studies. 
The third method ignores effects from the speed of sound and matches $\alpha_\theta$ to the bag model. The fourth method also uses the bag equation of state to furthermore simplify the prefactor, namely the approximation $D\theta/4e_+ \simeq \alpha_\theta/(\alpha_\theta+1)$. The fifth and sixth methods use $\alpha_p$ and $\alpha_e$ in the matching. We summarize all methods in Table~\ref{tab:methods}.
\begin{table}
\begin{center}
\begin{tabular}{ | c  | c |}
\hline
M1 & $K$ \\  
\hline
M2 & $\left(\frac{D\bar\theta}{4e_+}\right) \kappa_{\bar\theta}(\aNEW,c_s)\rvert_\nu$ \\
\hline
M3 & $\left(\frac{D\theta}{4e_+}\right) \kappa_\epsilon(\alpha_\theta)$ \\
\hline
M4 & $\left(\frac{\alpha_\theta}{\alpha_\theta+1}\right) \kappa_\epsilon(\alpha_\theta)$ \\
\hline
M5 & $\left(\frac{\alpha_p}{\alpha_p+1}\right) \kappa_\epsilon(\alpha_p)$ \\
\hline
M6 & $\left(\frac{\alpha_e}{\alpha_e+1}\right) \kappa_\epsilon(\alpha_e)$ \\
\hline
\end{tabular}
\caption{\small\label{tab:methods} Different methods to determine the efficiency $K$ by matching to the $\nu$-model or bag model.}
\end{center}
\end{table}

Finally, in Table~\ref{tab:final} we present the relative errors of the methods M2-M6 compared to the fully numerical result M1. The method M2 works better than expected with only a few percent deviation even for quite strong phase transitions. Methods M3 and M4 perform similarly with deviations up to $60\%$ which will correspond to an error of roughly $100\%$ in $\Omega_{GW}$. 
Notice that also the approximation $D\theta/4e_+ \simeq \alpha_\theta/(\alpha_\theta+1)$ contributes sizably to the discrepancy of M4.

Methods M5 and M6 perform quite poorly, as anticipated by our previous analysis.
Notice that M5 (using $\alpha_p$) underestimates the efficiency by a factor a few while M6 (using $\alpha_e$) overestimates by a factor few. Methods M3 and M4 are correct to $\mathcal{O}(1)$ and these methods have the advantage that a fit for $\kappa_\epsilon$ is readily available while $\kNEW$ is not. To remedy this issue, we provide a Python code in Appendix~\ref{app:kappa} that allows one to easily obtain $\kNEW$, such that M2 can be used without any additional effort.

\begin{table}
\begin{center}
\begin{tabular}{ | c  || c || c |c |c |c |c |}
\hline
model/method & M1 &M2 & M3 & M4 & M5 & M6\\  
\hline
SM$_1$ & 0.00143 & 0.45 \% & 4.99 \% & 3.55 \% & -88.45 \% & 713.34 \%  \\
\hline
SM$_2$ & 0.00401 & 0.43 \% & 1.70 \% & -0.72 \% & -66.69 \% & 351.90 \% \\
\hline
SM$_3$ & 0.00014 & 0.04 \% & 1.37 \% & 0.94 \% & -89.16 \% & 779.35 \% \\
\hline
SM$_4$ & 0.00039 & 0.04 \% & 0.42 \% & -0.32 \% & -67.85 \% & 405.11 \% \\
\hline
2step$_1$ & 0.00036 & -0.21 \% & 13.61 \% & 17.39 \% & -89.52 \% & 945.17 \% \\
\hline
2step$_2$ & 0.00563 & -0.80 \% & 15.68 \% & 21.90 \% & -50.01 \% & 366.20 \% \\
\hline
2step$_3$ & 0.00070 & -0.77 \% & 35.97 \% & 47.28 \% & -89.85 \% & 1235.34 \%\\
\hline
2step$_4$ & 0.01576 & -3.52 \% & 40.05 \% & 58.29 \% & -41.80 \% & 485.16 \% \\
\hline
\end{tabular}
\caption{\small\label{tab:final} Relative errors of the methods M2-M6 compared to the fully numerical result M1. The model parameters are given in Table~\ref{tab:SM} and~\ref{tab:2step} and a wall velocity of $\xi_w=0.9$ was used.}
\end{center}
\end{table}

\newpage
\section{Summary and conclusions}\label{sec:summary}

We studied the energy budget of cosmological phase transitions in a model-independent way. Our most important result is that for detonations the main parameters entering the determination of the efficiency coefficient are the speed of sound in the broken phase and a new strength parameter $\aNEW$  that we defined in equation~(\ref{eq:def:pt}). These parameters can be determined using the free energy of the system at the nucleation temperature. Parametrically, our approximation scheme works well as long as the temperature dependence of the speed of sound in the broken phase is not too large.

We determined the efficiency coefficient in a toy model that allows for a variable speed of sound in the broken phase which we call the $\nu$-model~\cite{Leitao:2014pda}. It turns out that the efficiency coefficient in realistic models can be obtained from this toy model by matching the strength parameter $\aNEW$ and the speed of sound to this toy model. The accuracy of this method for moderately strong phase transitions ($\aNEW \sim 0.15$) is a few percent. Methods that have been used previously in the literature only allow for an estimate that is of the right order of magnitude (if based on the trace of the energy-momentum tensor) or worse (if based on the pressure or energy difference between the two phases). In order to make our method readily available for phenomenological studies, we provide a Python code in the Appendix that determines the efficiency coefficient in the $\nu$-model.

One clear limitation of our approach is that so far it is only available for detonations. In case of deflagrations also the speed of sound in the symmetric phase will enter the fluid profile and has to be parametrized for high accuracy. In principle, one could generalize the $\nu$-model and introduce another parameter for the speed of sound in the symmetric phase. For the detonations, the result will not depend on this additional parameter (the relation~(\ref{eq:omatchnew}) is still exact in this model even though the expression for $\aNEW$ depends on the additional parameter), but for deflagrations it definitely does. The next best solution is to use the trace parameter $\alpha_\theta$ in the mapping to the bag equation of state which should give reasonable results. In any case, detonations are the most relevant case for GW production and the stability of hybrid solutions for strong phase transitions is still under debate~\cite{Huet:1992ex, Link:1992dm, Abney:1993ku, Ignatius:1993qn, KurkiSuonio:1995pp, KurkiSuonio:1995vy, Rezzolla:1996ey, Fragile:2003bw, Megevand:2013yua, Megevand:2014yua, Megevand:2014dua}, and we leave the analysis of deflagrations for the future.

In principle, our results can be applied right away to the common expressions for the GW power spectrum~\cite{Caprini:2015zlo, Caprini:2019egz} by just replacing
\be
K = \frac{\alpha \, \kappa}{\alpha+1} \to \left(\frac{D\bar\theta}{4e_+}\right) \, \kappa_{\bar\theta} (c_s, \aNEW, \xi_w) \, .
\ee
Furthermore, in simulations it is possible to measure the root mean square velocity $\bar U_f$ in the volume $\mathcal{V}$
\be
\bar U_f^2 = \frac{1}{w_+ \mathcal{V}} \int d\mathcal{V} \, \gamma^2 v^2 w \, ,
\ee
and the strength parameters $\alpha$ independently. Thus the relation $\bar U_f^2 = \ekin/w_+$ can be leveraged to test 
whether using only one spherical bubble in the calculation of $K$ is justified. This test was performed in~\cite{Hindmarsh:2015qta, Hindmarsh:2017gnf} assuming the bag model which gave deviations of up to~$\mathcal{O}(25\%)$.  In light of our analysis above, it would be helpful to measure the pseudotrace and the speed of sound in the broken phase along with the normal trace in hydrodynamic simulations and to reevaluate the comparison using the efficiency $\kappa_{\bar\theta}$. This would allow for the most accurate extrapolation of the gravitational wave power spectrum measured in large scale hydrodynamic simulations of the phase transitions~\cite{Hindmarsh:2015qta, Hindmarsh:2016lnk, Hindmarsh:2017gnf, Hindmarsh:2019phv} to arbitrary models. 

\section*{Acknowledgments}
We thank Ryusuke Jinno for useful discussions.
This work was funded by the Deutsche Forschungsgemeinschaft under Germany's Excellence Strategy - EXC 2121 ``Quantum Universe'' - 390833306.

\appendix
\section{The numerical efficiency coefficient in the $\nu$-model \label{app:kappa}}

\begin{figure}[t]
\begin{center}
\includegraphics[width=0.8\textwidth]{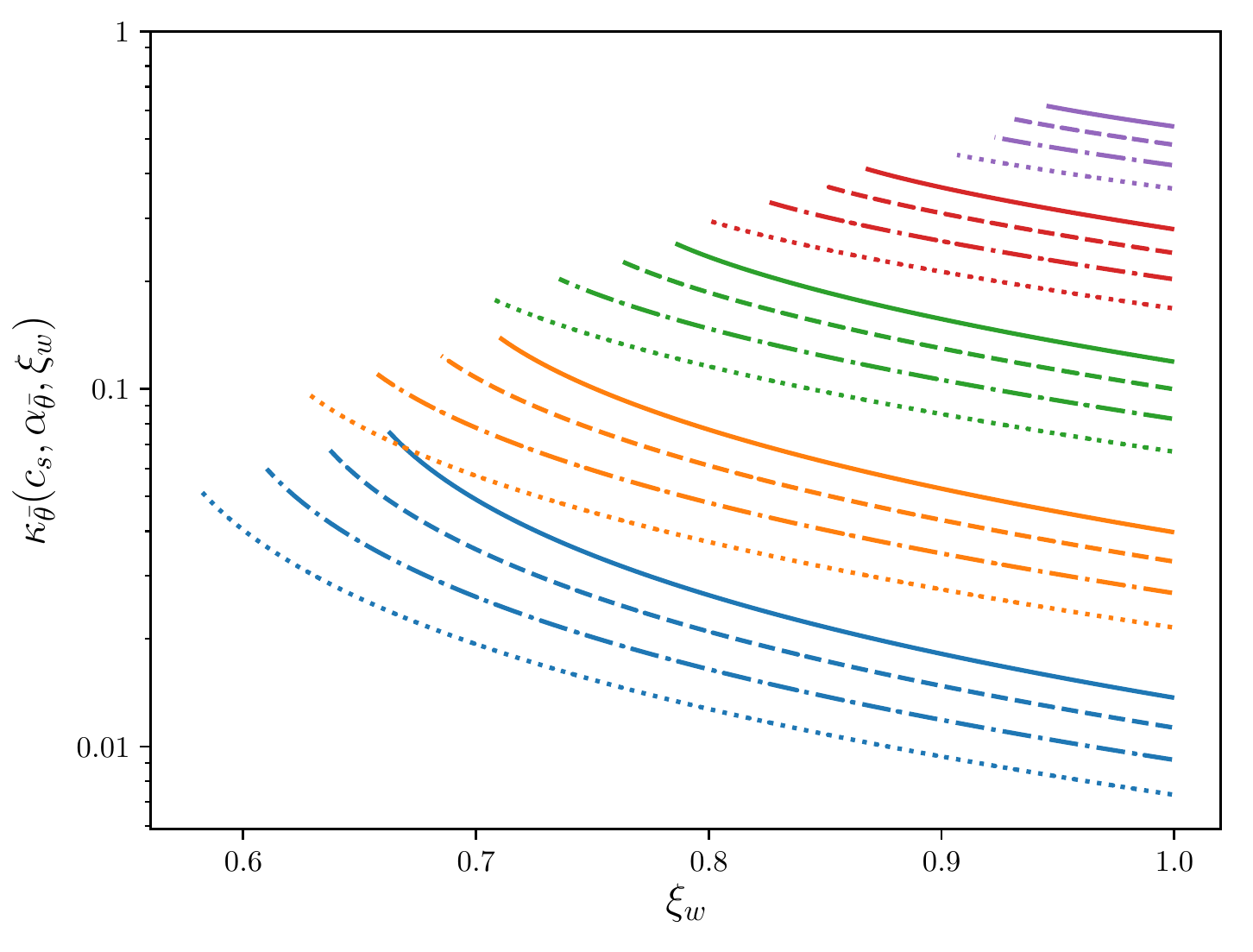}
\caption{\small Efficiency $\kappa_{\bar \theta}$ for the $\nu$-model as a function of the wall velocity $\xi_w$. The parameters are $\aNEW = [0.01, 0.03,0.1,0.3,1]$ from bottom to top with different colors and $c_s^2 = [12/36,11/36,10/36,9/36]$ from top to bottom with different line styles. 
\label{fig:kappasnu}}
\end{center}
\end{figure}

We have seen in section~\ref{sec:models} that using $\kNEW$ (method M2) significantly improves the accuracy compared to the methods that neglect the correction to the speed of sound (methods M3 and M4). Hence, a fit to $\kNEW$ is desirable. Unfortunately, an accurate fit is hard to produce since $\kNEW$ depends on three parameters ($\xi_w$, $\aNEW$ and $c_s$). 

However, the $\nu$-model is much simpler than a generic model due to the fact that the speed of sound is constant and many intermediate steps in the calculation of $\kNEW$ can be performed analytically. In particular, the fluid velocity behind the wall is given by
\bea
v_- &=& \frac{c + \sqrt{d}}{2(\nu-1) v_+} \, , \\
c &=& v_+^2 (\nu -1 + 3 \aNEW) + 1 - 3\aNEW \, , \\ 
d &=& c^2 - 4 v_+^2 (\nu-1) \, ,
\eea
and likewise the enthalpy behind the wall is given by
\be
\frac{w_-}{w_+} = \frac{ v_+/v_- (\nu - 1 + 3 \aNEW) - 1 + 3 \aNEW}{ \nu -1 - v_+/v_-} \, .
\ee
Furthermore, there are no detonations anymore below
\begin{equation}
	\xi_J=\frac{1+\sqrt{3\aNEW\left(1- c_s^2 +3 c_s^2 \aNEW\right)}}{1/c_s+3c_s\aNEW}.
\end{equation}
These are the strongest detonations possible for fixed wall velocity, the Jouguet-detonations.

We provide a Python code to calculate $\kNEW$ as a function of the speed of sound squared $c_s^2$, the strength of the phase transition $\aNEW$ and the wall velocity $\xi_w$  in Table~\ref{tab:code}.
Figure~\ref{fig:kappasnu} shows some numerical results for different values of $\aNEW$ and $c_s$. An increase in the strength obviously increases $\kNEW$ while an increase in the speed of sound $c_s$ does as well. The last fact is physically not obvious but can be traced back to the dependence of~(\ref{eq:vovNEW}) and~(\ref{eq:dvdxi}) on the speed of sound. 

\begin{table}
\begin{center}
\begin{tabular}{|l|l|}
\hline
\tt 01 & \tt import numpy as np \\
\tt 02 & \tt from scipy.integrate import odeint \\
\tt 03 & \tt from scipy.integrate import simps \\
\tt 04 & \tt  \\
\tt 05 & \tt def kappaNuModel(cs2,al,vp): \\
\tt 06 & \tt \quad nu = 1./cs2+1. \\
\tt 07 & \tt \quad tmp = 1.-3.*al+vp**2*(1./cs2+3.*al) \\
\tt 08 & \tt \quad disc = 4*vp**2*(1.-nu)+tmp**2 \\
\tt 09 & \tt \quad if disc<0: \\
\tt 10 & \tt \quad \quad print("vp too small for detonation") \\
\tt 11 & \tt \quad \quad return 0 \\
\tt 12 & \tt \quad vm = (tmp+np.sqrt(disc))/2/(nu-1.)/vp \\
\tt 13 & \tt \quad wm = (-1.+3.*al+(vp/vm)*(-1.+nu+3.*al)) \\
\tt 14 & \tt \quad wm /= (-1.+nu-vp/vm) \\
\tt 15 & \tt  \\
\tt 16 & \tt \quad def dfdv(xiw, v, nu): \\
\tt 17 & \tt \quad \quad xi, w = xiw \\
\tt 18 & \tt \quad \quad dxidv = (((xi-v)/(1.-xi*v))**2*(nu-1.)-1.) \\
\tt 19 & \tt \quad \quad dxidv *= (1.-v*xi)*xi/2./v/(1.-v**2)  \\
\tt 20 & \tt \quad \quad dwdv = nu*(xi-v)/(1.-xi*v)*w/(1.-v**2) \\
\tt 21 & \tt \quad \quad return [dxidv,dwdv] \\
\tt 22 & \tt  \\
\tt 23 & \tt \quad n = 501  \# change accuracy here \\
\tt 24 & \tt \quad vs = np.linspace((vp-vm)/(1.-vp*vm), 0, n)  \\
\tt 25 & \tt \quad sol = odeint(dfdv, [vp,1.], vs, args=(nu,)) \\
\tt 26 & \tt \quad xis, ws = (sol[:,0],-sol[:,1]*wm/al*4./vp**3) \\
\tt 27 & \tt  \\
\tt 28 & \tt \quad return simps(ws*(xis*vs)**2/(1.-vs**2), xis) \\
\hline
\end{tabular}
\caption{\small\label{tab:code}
Python code to calculate $\kNEW$ in the $\nu$-model as a function of the speed of sound squared $c_s^2$, the strength of the phase transition $\aNEW$ and the wall velocity $\xi_w$.}
\end{center}
\end{table}

\newpage

\bibliographystyle{JHEP}
\bibliography{BeyondBag}

\end{document}